\documentclass[aps,prl,preprint,nopacs,superscriptaddress]{revtex4}
\usepackage{graphicx}
\usepackage{verbatim}
\usepackage{mathrsfs}
\usepackage{amsmath,amsfonts,amssymb}
\usepackage{graphicx}

\def\3{2.8in}    
\def\2{2.5in}
\def\4{3.0in}

\def \beq {\begin{equation}}
\def \eeq {\end{equation}}
\begin{document}

\title{Theoretical Discovery/Prediction: Weyl Semimetal states in the TaAs material (TaAs, NbAs, NbP, TaP) class}

\author{Shin-Ming Huang\footnote{These authors contributed equally to this work.}}
\affiliation{Centre for Advanced 2D Materials and Graphene Research Centre National University of Singapore, 6 Science Drive 2, Singapore 117546}
\affiliation{Department of Physics, National University of Singapore, 2 Science Drive 3, Singapore 117542}
\author{Su-Yang Xu$^*$}
\email{suyangxu@princeton.edu}
\affiliation {Joseph Henry Laboratory, Department of Physics, Princeton University, Princeton, New Jersey 08544, USA}

\author{Ilya Belopolski$^*$}\affiliation {Joseph Henry Laboratory, Department of Physics, Princeton University, Princeton, New Jersey 08544, USA}

\author{Chi-Cheng Lee}
\affiliation{Centre for Advanced 2D Materials and Graphene Research Centre National University of Singapore, 6 Science Drive 2, Singapore 117546}
\affiliation{Department of Physics, National University of Singapore, 2 Science Drive 3, Singapore 117542}

\author{Guoqing Chang}
\affiliation{Centre for Advanced 2D Materials and Graphene Research Centre National University of Singapore, 6 Science Drive 2, Singapore 117546}
\affiliation{Department of Physics, National University of Singapore, 2 Science Drive 3, Singapore 117542}

\author{BaoKai Wang}
\affiliation{Centre for Advanced 2D Materials and Graphene Research Centre National University of Singapore, 6 Science Drive 2, Singapore 117546}
\affiliation{Department of Physics, National University of Singapore, 2 Science Drive 3, Singapore 117542}
\affiliation{Department of Physics, Northeastern University, Boston, Massachusetts 02115, USA}

\author{Nasser Alidoust}\affiliation {Joseph Henry Laboratory, Department of Physics, Princeton University, Princeton, New Jersey 08544, USA}
\author{Guang Bian}\affiliation {Joseph Henry Laboratory, Department of Physics, Princeton University, Princeton, New Jersey 08544, USA}
\author{Madhab Neupane}\affiliation {Joseph Henry Laboratory, Department of Physics, Princeton University, Princeton, New Jersey 08544, USA}


\author{Arun Bansil}
\affiliation{Department of Physics, Northeastern University, Boston, Massachusetts 02115, USA}

\author{Hsin Lin}
\email{nilnish@gmail.com}
\affiliation{Centre for Advanced 2D Materials and Graphene Research Centre National University of Singapore, 6 Science Drive 2, Singapore 117546}
\affiliation{Department of Physics, National University of Singapore, 2 Science Drive 3, Singapore 117542}

\author{M. Zahid Hasan}
\email{mzhasan@princeton.edu}
\affiliation {Joseph Henry Laboratory, Department of Physics, Princeton University, Princeton, New Jersey 08544, USA}


\begin{abstract}
The recent discoveries of Dirac fermions in graphene and on the surface of topological insulators have ignited worldwide interest in physics and materials science. A Weyl semimetal is an unusual crystal where electrons also behave as massless quasi-particles but interestingly they are not Dirac fermions. These massless particles, Weyl fermions, were originally considered in massless quantum electrodynamics but have not been observed as a fundamental particle in nature. A Weyl semimetal provides a condensed matter realization of Weyl fermions, leading to unique transport properties with novel device applications. Such a semimetal is also a topologically non-trivial metallic phase of matter extending the classification of topological phases beyond insulators. The signature of a Weyl semimetal in real materials is the existence of unusual Fermi arc surface states, which can be viewed as half of a surface Dirac cone in a topological insulator. Here, we identify the first Weyl semimetal in a class of stoichiometric materials, which break crystalline inversion symmetry, including TaAs, TaP, NbAs and NbP. Our first-principles calculations on TaAs reveal the spin-polarized Weyl cones and Fermi arc surface states in this compound. We also observe pairs of Weyl points with the same chiral charge which project onto the same point in the surface Brillouin zone, giving rise to multiple Fermi arcs connecting to a given Weyl point. Our results show that TaAs is the first topological semimetal identified which does not depend on fine-tuning of chemical composition or magnetic order, greatly facilitating an exploration of Weyl physics in real materials.
(Note added: This theoretical prediction of November 2014 (see paper in Nature Communications at http://www.nature.com/ncomms/2015/150612/ncomms8373/full/ncomms8373.html ) was the basis for the first experimental discovery of Weyl Fermions and topological Fermi arcs in TaAs recently published in Science (2015) at http://www.sciencemag.org/content/early/2015/07/15/science.aaa9297.abstract )

\end{abstract}
\date{\today}
\maketitle

The rich correspondence between high energy and condensed matter physics has led to a deeper understanding of spontaneous symmetry breaking, phase transitions, renormalization and many other fundamental phenomena in nature, with important consequences for practical applications using magnets, superconductors and other novel materials \cite{Weyl, Wilczek, Volovik2003}. Recently, there has been considerable progress in realizing particles previously considered in high energy physics as emergent quasiparticle excitations of crystalline solids, such as Majorana fermions and Dirac fermions \cite{Ashvin_Review, Haldane, Balents_viewpoint, MF, Graphene, Hasan2010, Qi2011, NaBi, CdAs}. Materials that host these exotic particles exhibit unique properties and hold promise for applications such as fault-tolerant quantum computation, low-power electronics and spintronics. Weyl fermions were originally considered in massless quantum electrodynamics, but have not been observed as a fundamental particle in nature \cite{Weyl}. Recently, it was theoretically understood that Weyl fermions can arise in some novel semimetals with nontrivial topology \cite{Ashvin_Review, Haldane, Balents_viewpoint, Murakami2007}. A Weyl semimetal has an electron band structure with singly-degenerate bands that have bulk band crossings, Weyl points, with a linear dispersion relation in all three momentum space directions moving away from the Weyl point. These materials can be viewed as an exotic spin-polarized, three-dimensional version of graphene. However, unlike the two-dimensional Dirac cones in graphene \cite{Graphene}, the three-dimensional Dirac cones in Na$_3$Bi and Cd$_3$As$_2$ \cite{NaBi, CdAs} or the two-dimensional Dirac cone surface states of Bi$_2$Se$_3$ \cite{Hasan2010, Qi2011}, the degeneracy associated with a Weyl point depends only on the translation symmetry of the crystal lattice. This makes the unique properties associated with this electron band structure more robust. Moreover, due to its nontrivial topology, a Weyl semimetal can exhibit novel Fermi arc surface states. Previously, Fermi arcs were only found in high-$T_\textrm{c}$ superconductors due to strong electron correlation effects. Recent theoretical advances in topological physics have now made it possible to realize Fermi arc surface states in a weakly interacting semimetal. Both the Weyl fermions in the bulk and the Fermi arc states on the surface of a Weyl semimetal are predicted to show unusual transport phenomena, which can be used in future device applications. Weyl fermions in the bulk can give rise to negative magnetoresistance, the quantum anomalous Hall effect, non-local transport and local non-conservation of ordinary current \cite{Chiral, Chiral_Qi, Qi_review, Nonlocal}. Fermi arc states on the surface are predicted to show novel quantum oscillations in magneto-transport and quantum interference effects in tunneling spectroscopy \cite{Ojanen, Ashvin2, Hosor}. Because of the fundamental and practical interest in Weyl semimetals, it is crucial that robust candidate materials be found.

It is theoretically known that a Weyl semimetal can only arise in a crystal where time-reversal symmetry or inversion symmetry is broken. A number of magnetically ordered or inversion symmetry breaking materials have been proposed as Weyl semimetal candidates \cite{Wan2011, Burkov2011, HgCrSe, HgCdTe, Vanderbilt}. However, despite extensive effort in experiment, a Weyl semimetal has yet to be realized in any of the compounds proposed thus far. One concern is that the existing proposals require either magnetic ordering in sufficiently large domains \cite{Wan2011, Burkov2011, HgCrSe, HgCdTe} or fine-tuning of the chemical composition to within 5\% in an alloy \cite{Burkov2011, HgCdTe, Vanderbilt}, which are demanding in real experiments. Here we propose the first Weyl semimetal in a stoichiometric, inversion-breaking, single-crystal material, TaAs. Unlike previous predictions, our proposal does not depend on magnetic ordering over sufficiently large domains, because our material relies on inversion symmetry breaking rather than time-reversal symmetry breaking. The compound we propose is also stoichiometric and does not depend on fine-tuning chemical composition in an alloy. Single crystals of TaAs have been grown \cite{TaAs_Crystal}. We believe that this material is promising for experimental realization for the Weyl semimetal phase in this compound.

\begin{figure*}
\centering
\includegraphics[width=17cm]{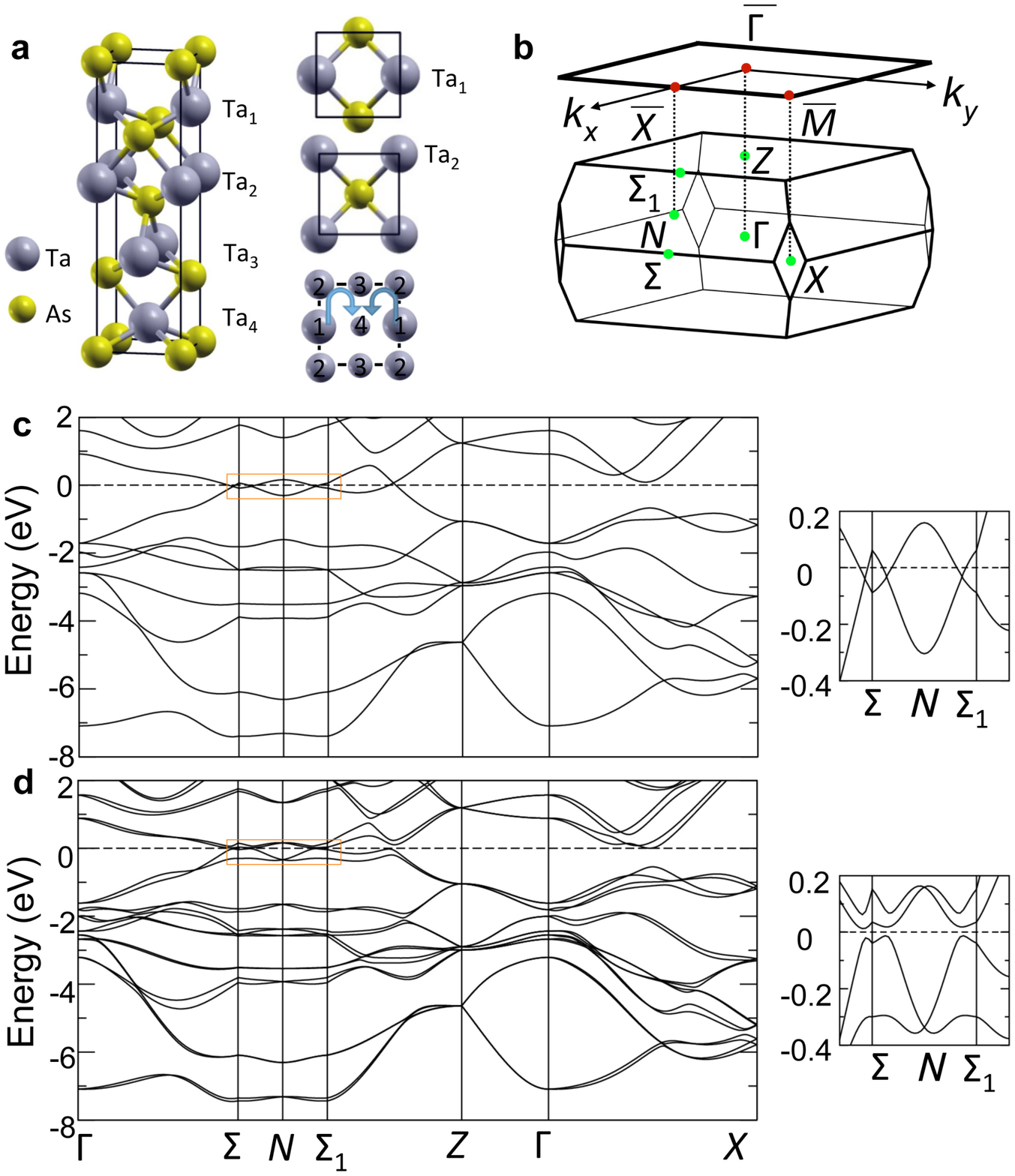}
\caption{\label{Fig1}\textbf{The crystal and electronic structure of the Weyl semimetal candidate TaAs.} \textbf{a,} Boby-centred tetragonal structure of TaAs, shown as stacked TaAs layers. An electric polarization is induced due to dimples in the TaAs lattice. The arrangement of Ta atoms for each layer (Ta$_1$ to Ta$_4$) is illustrated in the bottom-right panel. \textbf{b,} The bulk and surface Brillouin zones. \textbf{c,} The bulk electronic structure of TaAs in the absence of spin-orbit coupling from DFT.  \textbf{d,} The same as panel (\textbf{c}) but in the presence of spin-orbit coupling. The right panels of \textbf{c,d} show the band structure in the vicinity of the Fermi level along the $\Sigma-N-\Sigma _{1}$ direction.}
\end{figure*}

Tantalum arsenide, TaAs, crystalizes in a body-centered tetragonal lattice system (Fig.~\ref{Fig1}\textbf{a}). The lattice constants are $a=3.437$ \r{A} and $c=11.646$ \r{A}, and the space group is I4$_{1}$md (\#109, $C_{4v}$). The crystal consists of interpenetrating Ta and As sub-lattices, where the two sub-lattices are shifted by $\left( \frac{a}{2},\frac{a}{2},\delta \right) $, $\delta \approx \frac{c}{12}$. There are two Ta atoms and two As atoms in each primitive unit cell. It is important to note that the system lacks a horizontal mirror plane and thus inversion symmetry. This makes it possible to realize an inversion breaking Weyl semimetal in TaAs. We also note that the $C_4$ rotational symmetry is broken at the (001) surface because the system is only invariant under a four-fold rotation with a translation along the out-of-plane direction. The bulk and (001) surface Brillouin zones (BZ) are shown in Fig.~~\ref{Fig1}\textbf{b}, where high symmetry points are also noted.

The ionic model would suggest that the Ta and As atoms are in the $3^{+}$ and $3^{-}$ valence states, respectively. This indicates that the lowest valence band arises from $4p$ electrons in As and $5d^{2}$ electrons in Ta, whereas the lowest conduction band primarily consists of $5d$ electrons in Ta. However, we may expect Ta $5d$ electrons to have a broad bandwidth because of the wide extent of the atomic orbitals. This leads to strong hybridization with the As $4p$ states, which may suggest that the conduction and valence bands are not entirely separated in energy and have a small overlap, giving rise to a semimetal. In Fig.~\ref{Fig1}\textbf{c}, we present the bulk band structure in the absence of spin-orbit coupling. The conduction and valence bands cross each other along the $\Sigma-N-\Sigma_1$ trajectory, which further indicates that TaAs is a semimetal. In the presence of spin-orbit coupling, the band structure is fully gapped along the high symmetry directions considered in Fig.~\ref{Fig1}\textbf{d}. However, Weyl points which are shifted away from the high symmetry lines arise after spin-orbit coupling is taken into account. Below, we consider the Weyl points in the bulk BZ. We also note that the double degeneracy of bands is lifted in the presence of spin-orbit coupling except at the Kramers' points, which confirms that TaAs breaks inversion symmetry but respects time-reversal symmetry.

\begin{figure}
\includegraphics[width=17cm]{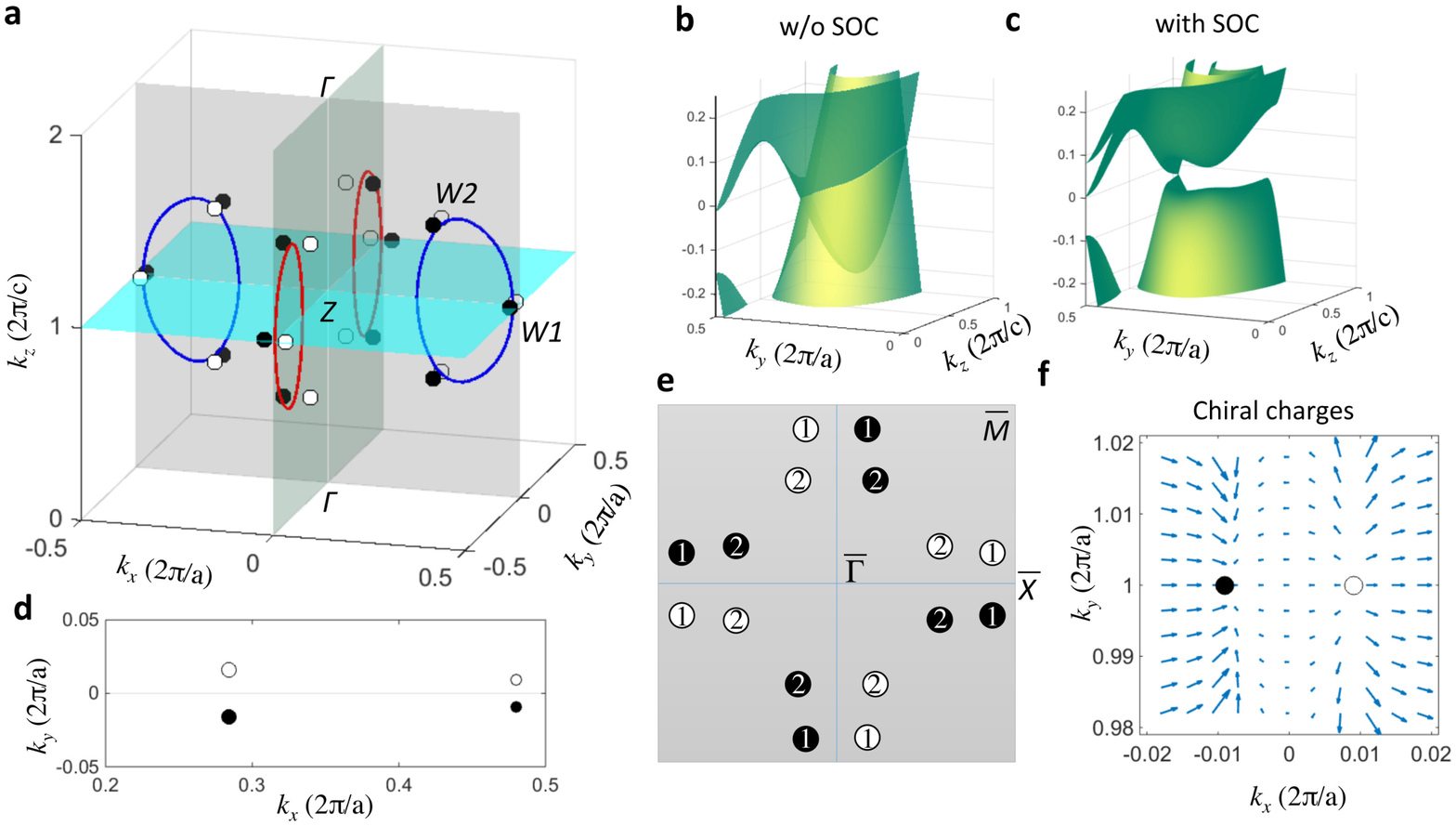}
\caption{\label{Fig2}\textbf{Weyl points and chiral charges in TaAs.} \textbf{a,} In the absence of spin-orbit coupling, there are two line nodes on the $k_{x}=0$ mirror plane, $M_x$, and two line nodes on the $k_{y}=0$ mirror plane, $M_y$. In the presence of spin-orbit coupling, each line node vaporizes into six Weyl points. The Weyl points are denoted by small circles. Black and white show the opposite chiral charges of the Weyl points. \textbf{b,} The band structure around one of the red line nodes. \textbf{c,} The band structure near one of the Weyl points at $k_xa=0.01\pi$. The Weyl point is shifted away from the $k_{x}=0$ mirror plane. \textbf{d,} The projection of the bulk Weyl points around the $\bar{\Gamma}-\bar{X}$ line. \textbf{e,} A schematic for the projection of all the Weyl points on the (001) surface Brillouin zone. We denote the eight Weyl points that are located on the $k_z=2\pi/c$ plane as $W_1$ and the other sixteen Weyl points as $W_2$. \textbf{f,} Calculated spin texture $(S_{y},S_{z})$ in the vicinity of two Weyl points with opposite chiral charges.}
\end{figure}

To better understand the Weyl points in TaAs, we first examine the band crossings near the Fermi level in the absence of spin-orbit coupling. These band crossings take the form of a closed curve, or line node, on the $M_x$ and $M_y$ mirror planes of the BZ, shown in red and blue in Fig.~\ref{Fig2}\textbf{a}. We present the band structure near one of the red nodal lines, on the $M_x$ mirror plane, in Fig.~\ref{Fig2}\textbf{b}. The conduction and valence bands dip into each other, giving rise to a line node crossing. Next, we include spin-orbit coupling. This causes each line node to vaporize into six Weyl points shifted slightly away from the mirror planes, shown as small circles in Fig.~\ref{Fig2}\textbf{a}. There are 24 Weyl points in total: 8 Weyl points on the $k_z = 2\pi/c$ plane, which we call $W_1$, and 16 Weyl points away from the $k_z = 2\pi/c$ plane, which we call $W_2$. We present the band structure near one of the Weyl points $W_1$ in Fig.~\ref{Fig2}\textbf{c}, where a point band touching is clearly observed. Next, we consider how the Weyl points project on the (001) surface BZ. We show a small region in the surface BZ around the $\bar{\Gamma}-\bar{X}$ line, which includes the projections of six Weyl points, two from $W_1$ and four from $W_2$. A schematic of the projections of all 24 Weyl points on the surface BZ is shown in Fig.~\ref{Fig2}\textbf{e}. We find that for all points $W_2$, two Weyl points project on the same point in the surface BZ. We indicate the number of Weyl points that project on a given point in the surface BZ in Fig.~\ref{Fig2}\textbf{e}. Lastly, we study the chirality of the Weyl points by calculating the Berry flux through a closed surface enclosing a Weyl point. We call positive the chiral charges which are a source of Berry flux and negative the chiral charges which are a sink of Berry flux. We color the positive chiral charges white and the negative chiral charges black. In Fig.~\ref{Fig2}\textbf{f}, we show the spin texture of the band structure near two Weyl points $W_1$. In this case, the spin texture is found to be consistent with the texture of the Berry curvature. Moreover, it turns out that the points $W_2$ project on the surface BZ in pairs which carry the same chiral charge.

\begin{figure}
\centering
\includegraphics[width=17cm]{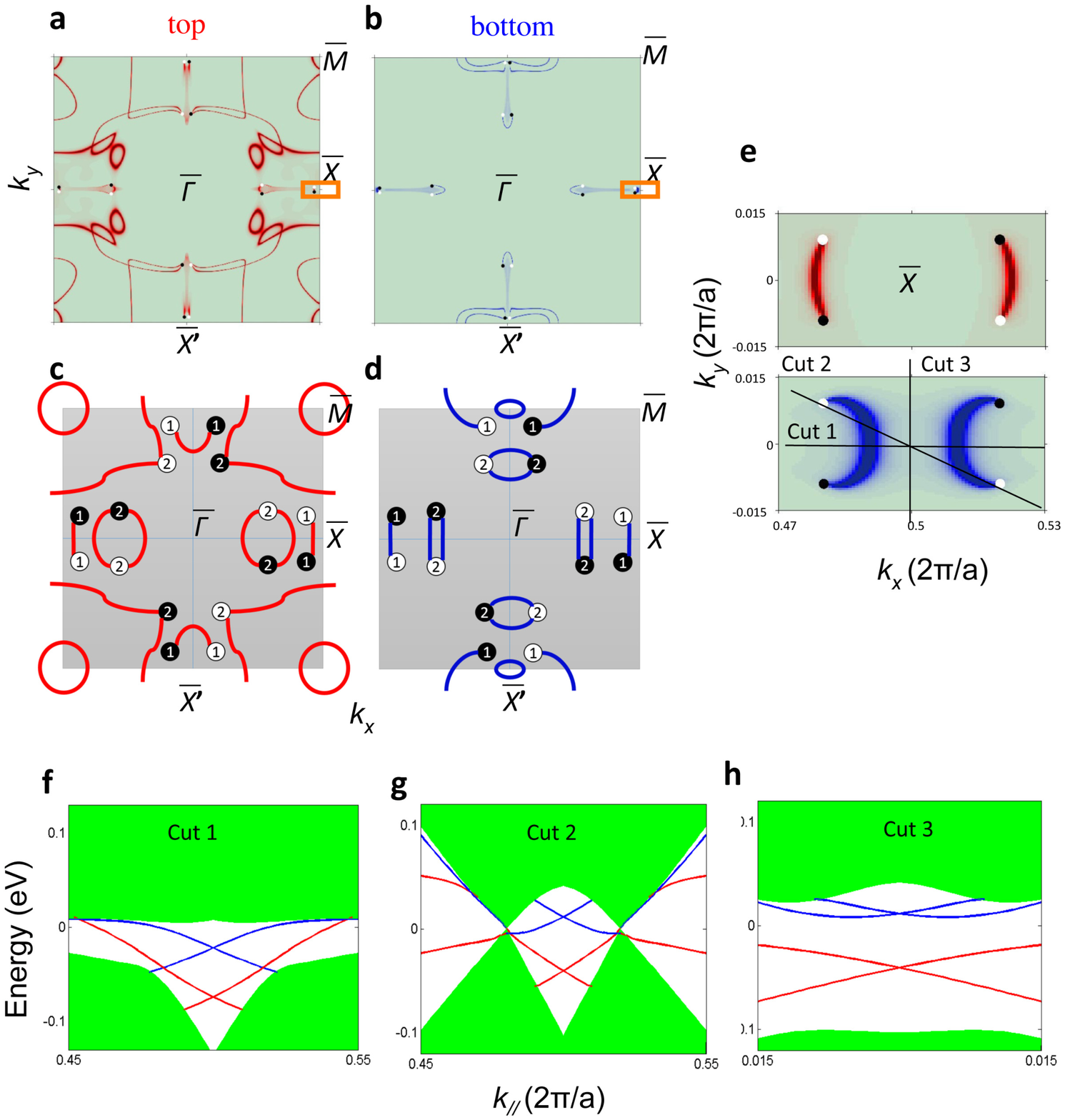}
\caption{\label{Fig3} \textbf{Fermi arc surface states in TaAs.}}
\end{figure}

Another key signature of a Weyl semimetal is the presence of Fermi arc surface states which connect the Weyl points in pairs in the surface BZ. We present calculations of the (001) surface states in Fig.~\ref{Fig3}. We show the surface states on the top surface in Fig.~\ref{Fig3}\textbf{a} and the bottom surface in Fig.~\ref{Fig3}\textbf{b}. We find surface Fermi arcs that connect Weyl points of opposite chirality in pairs. To better understand the rich structure of the Fermi arcs, we show a schematic of the surface states on the top surface in Fig.~\ref{Fig3}\textbf{c} and the bottom surface in Fig.~\ref{Fig3}\textbf{d}. Note that one Fermi arc connects each pair of points $W_1$. However, two Fermi arcs connect to each projection of points $W_2$, because they project in pairs with the same chiral charge, as discussed above. This leads to Fermi arcs which connect the points $W_2$ in a closed loop of surface states. The largest Fermi arc loop on the top surface threads through four projected Weyl points in the surface BZ. We also note that the Fermi surface of the surface states from the top is very different from that of the bottom (Figs.~\ref{Fig3}\textbf{a,b}), consistent with broken inversion symmetry in this system. In addition, because the presence of a surface breaks the $C_4$ screw symmetry, the surface states are also very different along the $k_x$ and $k_y$ directions of the surface BZ. Finally, we observe closed Fermi surfaces which do not intersect Weyl points and do not form Fermi arcs. These extra Fermi surfaces reflect how different ways of annihilating the Weyl points would give rise to an insulator with a different topological invariant (see the discussion below). We present a particularly simple set of Fermi arcs arising near the $\bar{X}$ point in Fig.~\ref{Fig3}\textbf{e}, including surface states from both the top and bottom surfaces. To visualize the arc nature of the surface states, we present three energy dispersion cuts along the directions indicated in Fig.~\ref{Fig3}\textbf{e}. Along Cut 1, shown in Fig.~\ref{Fig3}\textbf{f}, we see a Dirac cone connecting the bulk valence and conduction bands across the bulk band gap, exactly like a topological insulator. Along Cut 2, shown in Fig.~\ref{Fig3}\textbf{g}, we see the projected bulk Weyl cones, with surface states which pass through the Weyl points. Lastly, in Cut 3, shown in Fig.~\ref{Fig3}\textbf{h}, we observe a full band gap. The surface states along this cut are trivial because they do not connect across the bulk band gap.

\section{Discussion}

A number of candidates for a Weyl semimetal have been previously proposed. However, existing proposals have proven difficult to carry out because they rely on magnetic ordering to break time-reversal symmetry or fine-tuning the chemical composition of an alloy. Magnetic order can be difficult to predict from first-principles, may be difficult to measure experimentally, and most importantly, may not form large enough domains in a real sample for the properties of the Weyl semimetal to be preserved. Fine-tuning chemical composition is typically challenging to achieve and introduces disorder, limiting the quality of single crystals. For instance, a proposal for a Weyl semimetal in Y$_2$Ir$_2$O$_7$ \cite{Wan2011} assumes an all-in, all-out magnetic order, which is challenging to verify in experiment \cite{Fisher, YHZhang}. Another proposed Weyl semimetal, HgCr$_2$Se$_4$ \cite{HgCrSe}, has a clear ferromagnetic order, but because of the cubic structure there is no preferred magnetization axis, likely leading to the formation of many small ferromagnetic domains. Moreover, a proposal in Hg$_{1-x-y}$Cd$_x$Mn$_y$Te \cite{HgCdTe} requires straining the sample to break cubic symmetry, applying an external magnetic field and fine-tuning the chemical composition. Another proposal for the inversion breaking compounds LaBi$_{1-x}$Sb$_x$Te$_3$ and LaBi$_{1-x}$Sb$_x$Te$_3$ \cite{Vanderbilt} requires fine-tuning the chemical composition to within 5\%. Despite extensive experimental effort, a Weyl semimetal has not been realized in any of these compounds. We believe that in large part the difficulty stems from relying on magnetic order to break time-reversal symmetry and fine-tuning the chemical composition to achieve the desired band structure. We note that, in contrast to time-reversal symmetry breaking systems, large single crystals of inversion symmetry breaking compounds exist, such as the large bulk Rashba material BiTeI, where the crystal domains are sufficiently large so that the Rashba band structure is clearly observed in photoemission spectroscopy \cite{BiTeI}. We propose that TaAs overcomes the difficulties of previous candidates because it realizes a Weyl semimetal in a stoichiometric, inversion-breaking, single-crystal material.

We have also studied other members of this class of compounds, TaP, NbAs and NbP. They have the same crystal structure as TaAs. Our results show that their band structure is qualitatively the same as that of TaAs, in that the conduction and valence bands form line nodes in the absence of spin-orbit coupling and a gap is opened in the presence of spin-orbit coupling, leading to a Weyl semimetal. In particular, since Nb and P have quite weak spin-orbit coupling, our calculations suggest that NbP is better described as a topological nodal-line semimetal. Thus NbP might offer the possibility to realize novel nodal line band crossings in the bulk and ``drumhead'' surface states stretching across the nodal line on the surface \cite{Burkov}.
	
Next, we provide some general comments on the nature of the phase transition between a trivial insulator, a Weyl semimetal and a topological insulator, illustrated in Fig.~\ref{Fig4}\textbf{a}. When a system with inversion symmetry undergoes a topological phase transition between a trivial insulator and a topological insulator, the band gap necessarily closes at a Kramers' point. If we imagine moving the system through the phase transition by tuning a parameter $m$, then there will be a critical point where the system is gapless. In a system which breaks inversion symmetry, there will instead be a finite range of $m$ where the system remains gapless, giving rise to a Weyl semimetal phase. In this way, the Weyl semimetal phase can be viewed as an intermediate phase between a trivial insulator and a topological insulator, where the bulk band gap of a trivial insulator closes and Weyl points of opposite chiral charge nucleate from the bulk band touchings. As $m$ is varied, the Weyl points thread surface states through the surface BZ and eventually annihilate each other, allowing the bulk band gap to reopen with a complete set of surface states, giving rise to a topological insulator. This understanding of a Weyl semimetal as an intermediate phase between a trivial insulator and a topological insulator offers some insight into the closed Fermi surfaces we find in TaAs around the $\bar{M}$ point of the top surface and the $\bar{X}'$ point of the bottom surface. We propose that these surface states reflect the topological invariant we would find if we annihilated the Weyl points to produce a bulk insulator. We consider the bottom surface, and annihilate the Weyl points in pairs in the obvious way to remove all surface states along $\bar{\Gamma}-\bar{X}$. Then, we can annihilate the remaining Weyl points to produce two concentric Fermi surfaces around the $\bar{X}'$ point. Since this is an even number of surface states, we find that this way of annihilating the Weyl points gives rise to a trivial insulator. If, instead, we annihilate the remaining Weyl points to remove the Fermi arc connecting them, only the closed Fermi surface would be left around $\bar{X}'$, giving rise to a topological insulator. A similar analysis applies to the top surface.

Finally, we highlight an interesting phenomenon regarding how an electron travels along a constant energy contour in our predicted Weyl semimetal, TaAs. Fig.~\ref{Fig4}\textbf{b} shows the calculated Fermi surface contours on both the top and the bottom surfaces near the $\bar{X}$ point. We consider an electron initially occupying a state in one of the red Fermi arcs, whose wavefunction is therefore localized on the top surface in real space. We ask how the wavefunction evolves as the electron traces out the constant energy contour. As the electron moves along the Fermi arc (the red arc on the left-hand side of Fig.~\ref{Fig4}\textbf{c}), it will eventually reach a Weyl point, where its wavefunction will unravel into the bulk. The electron will then move in real space to the bottom surface of the sample. Then it will follow a Fermi arc on that surface (the blue arc on the left-hand side of Fig.~\ref{Fig4}\textbf{b}), and reach another Weyl point. The electron will next travel through the bulk again and reach the top surface, returning to the same Fermi arc where it began. This novel trajectory is predicted to show exotic surface transport behavior as proposed in Ref. \cite{Ashvin2}. This is one example of the unique transport phenomena predicted in Weyl semimetals. To study these effects both from fundamental interest and for novel devices, here we provides a promising candidate for a Weyl semimetal in a stoichiometric, inversion-breaking, single-crystal material, as represented by TaAs. We hope that our theoretical prediction will soon lead to the experimental discovery of the first Weyl semimetal in nature.

\bigskip
\bigskip
\textbf{Methods}
\newline
First-principles calculations were performed by OPENMX code based on norm-conserving pseudopotentials generated with multi-reference energies and optimized pseudoatomic basis functions within the framework of the generalized gradient approximation (GGA) of density functional theory (DFT) \cite{Perdew, Ozaki}. Spin-orbit coupling was incorporated through $j$-dependent pseudo-potentials \cite{Theurich}. For each Ta atom, three, two, two, and one optimized radial functions were allocated for the $s$, $p$, $d$, and $f$ orbitals ($s3p2d2f1$), respectively, with a cutoff radius of 7 Bohr. For each As atom, $s3p3d3f2$ was adopted with a cutoff radius of 9 Bohr. A regular mesh of 1000 Ry in real space was used for the numerical integrations and for the solution of the Poisson equation. A $k$ point mesh of ($17 \times 17 \times 5$) for the conventional unit cell was used and experimental lattice parameters were adopted in the calculations. Symmetry-respecting Wannier functions for the As $p$ and Ta $d$ orbitals were constructed without performing the procedure for maximizing localization and a real-space tight-binding Hamiltonian was obtained \cite{Weng}. This Wannier function based tight-binding model was used to obtain the surface states by constructing a slab with 80-atomic-layer thickness with Ta on the top and As on the bottom.

\newpage

\clearpage
\clearpage

\clearpage
\addtocounter{figure}{-1}
\begin{figure*}[t!]
\caption{\textbf{Fermi arc surface states in TaAs.} \textbf{a,} The (001) surface states on the top surface of TaAs. The Weyl points are denoted by small black and white circles. \textbf{b,} The same as panel (\textbf{a}) but on the bottom surface. \textbf{c,d,} Schematics of the surface Fermi surfaces for the top and the bottom surfaces. \textbf{e,} A close-up of the band structure on both the top and the bottom surfaces near the $\bar{X}$ point, as indicated by the orange squares in panels (\textbf{a,b}). \textbf{f-h,} Energy dispersions of the electronic structure along three momentum space directions, as noted in panel (\textbf{e}).}
\end{figure*}

\clearpage
\begin{figure}
\centering
\includegraphics[width=17cm]{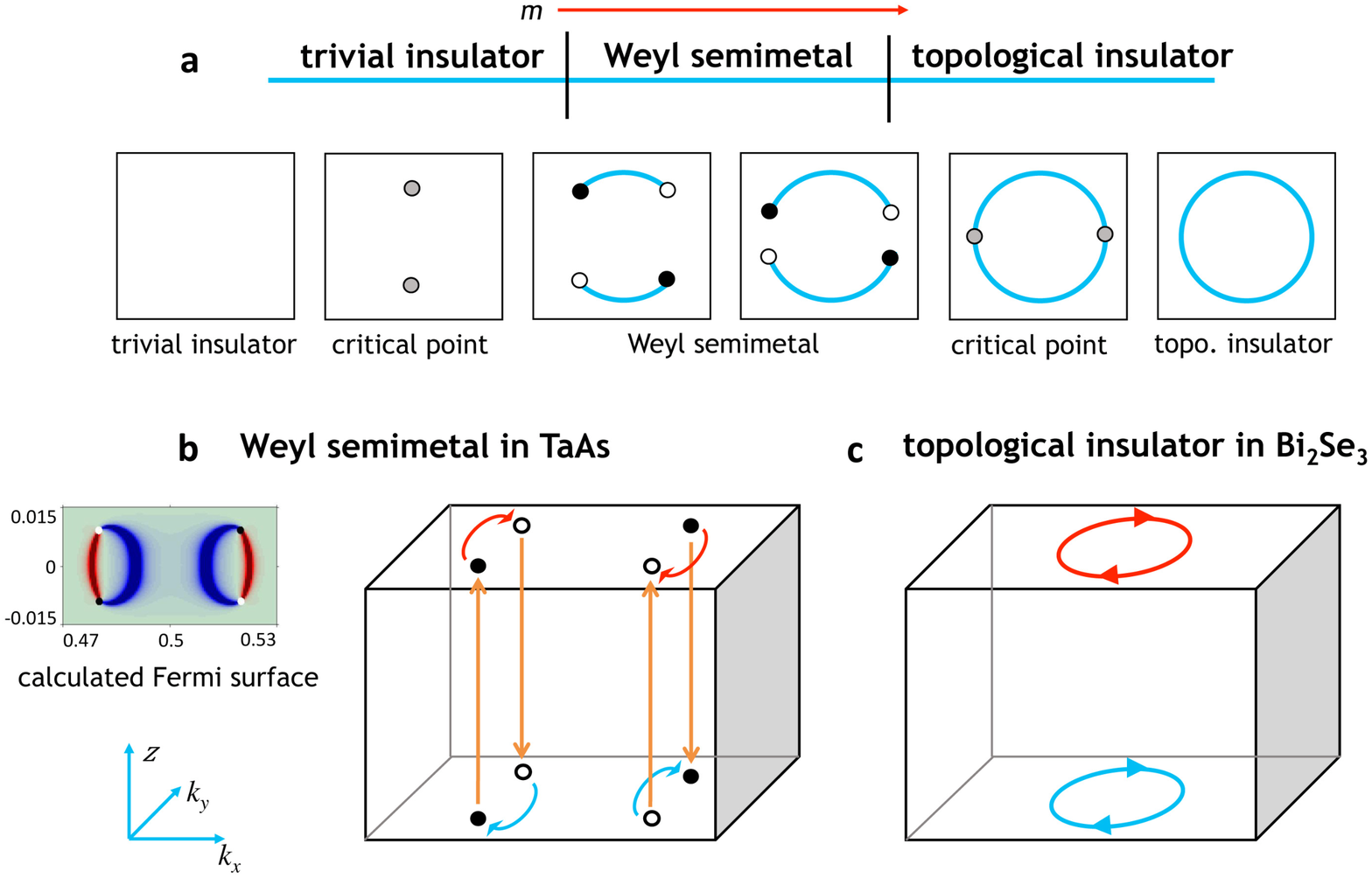}
\caption{\label{Fig4} \textbf{Schematic illustrations of several important aspects of an inversion breaking Weyl semimetal.} \textbf{a,} A Weyl semimetal can be understood as an intermediate phase between a trivial insulator and a topological insulator. Consider the evolution of Weyl points and surface states as a function of a tuning parameter $m$. In the trivial insulator phase, there are no surface states. At a critical $m$, the bulk conduction and valence bands touch each other. For higher values of $m$, each bulk band touching splits into two Weyl points with opposite chiral charges, and a Fermi arc surface state connects them. In this regime, the system is a Weyl semimetal. As $m$ increases further, eventually the Weyl points pair up again, giving rise to another critical point. For still higher values of $m$, the Weyl points annihilate and a full bulk band gap reopens, leaving a gapless surface state. The system is then a topological insulator. \textbf{b,} The calculated Fermi surface of TaAs near the $\bar{X}$ point. \textbf{c,} An electron tracing out a constant energy contour on a Fermi arc will follow an unusual path in real space. Initially, the electron wavefunction will be localized near }
\end{figure}
\addtocounter{figure}{-1}
\begin{figure*}[t!]
\caption{the top surface of the sample. As the electron proceeds around the Fermi arc, it will eventually approach a Weyl point and the wavefunction will unravel into the bulk. The electron will then move in real space to the bottom surface of the sample, and there it will follow the Fermi arc on that surface. The electron will eventually approach another Weyl point, pass again through the bulk and remerge in the same Fermi arc where it began, completing the circuit. \textbf{d,} By contrast, a topological insulator has a full bulk band gap, so an electron tracing out a constant energy contour in the surface states will not encounter a bulk state and the wavefunction will always remain localized on the same surface.}
\end{figure*}

\end{document}